\documentclass[twocolumn,amsmath,amssymb,aps,prl,floatfix,superscriptaddress]{revtex4}
\usepackage{graphicx} 
\usepackage{dcolumn} 
\usepackage{bm} 
\usepackage{epsfig}
\usepackage{color}
\usepackage{multirow}

\usepackage{amsmath}

\begin{document}
\title{Formal Valence, $d$ Occupation, and Charge-Order Transitions}

\author{Yundi Quan}
\affiliation{Physics Department, University of California Davis}

\author{Victor Pardo}
\affiliation{Physics Department, University of California Davis}
\affiliation{Departamento de F\'{i}sica Aplicada, Universidade Santiago de Compostela, Spain}

\author{Warren E. Pickett}
\email{wepickett@ucdavis.edu}
\affiliation{Physics Department, University of California Davis}

\begin{abstract}
While the formal valence and charge state concepts have been tremendously important in materials
physics and chemistry, their very loose connection to actual charge leads to uncertainties
in modeling behavior and interpreting data. We point
out, taking several transition metal oxides (La$_2$VCuO$_6$, YNiO$_3$,
CaFeO$_3$, AgNiO$_2$, V$_4$O$_7$) as examples, that while dividing the crystal
charge into atomic contributions is an ill-posed activity, the $3d$ {\it occupation}
of a cation (and more particularly, differences) is readily available 
in first principles calculations.
We discuss these examples, which include distinct charge states and charge-order
(or disproportionation)
systems, where different ``charge states'' of cations have {\it identical}
$3d$ orbital occupation.  Implications for theoretical modeling of such
charge states and charge-ordering mechanisms are discussed.
\end{abstract}
\maketitle


Spin ordering, and often orbital ordering, is normally unambiguous, as these
properties are subject to direct observation by magnetic and spectroscopic measurements,
respectively.  Charge ordering (CO) and the actual
charge of an ion is rarely measured directly, and the formal charge of an
ion in the solid state can be a point of confusion and contention.
Valence, oxidation number, and formal charge are concepts borrowed from chemistry,
where it is emphasized they do not represent actual charge\cite{chem1,chem2} and
have even been labeled hypothetical.\cite{chem1} 
As the interplay between spin, charge, orbital, and lattice degrees of freedom become more
closely watched\cite{DIKGAS} and acknowledged to be a complex phenomenon, 
disproportionation and CO have become entrenched as the explanation
of several high profile metal-insulator transitions (MIT). 
The possibility that CO in the charge transfer regime is associated with the oxygen
sublattice, with negligible participation of the metal, has been raised\cite{mizokawa}
and considered as an alternative.\cite{Mazin}

Charge density is a physical observable of condensed 
matter, and the desire to assign charge to atoms has evident pedagogical
value, so theoretical approaches
have been devised to share it amongst constituent nuclei.  Mulliken charge population, 
which socializes shared charge (divides it evenly between overlapping orbitals) is
notoriously sensitive to the local orbital basis set that is required to specify it.
Born effective charges are dynamical properties
and are often quite different from any conceivable formal charge or actual charge.
Integrations over various volumes have been used a great deal, but dividing the static 
crystal charge density into atomic contributions is, undeniably, 
an ill-defined activity.

A possibility that has not been utilized is that,
taking $3d$ oxides as an example, there is
a directly relevant metric that is well defined: the $d$ occupation $n_d$.
This quantity is in fact what the physical
picture of formal charge or oxidation state brings to mind. 
$3d$ cations, in their various environments and charge states, have maxima in their 
spherically averaged radial density
$\bar{\rho}(r)$ 
in the range 0.6-0.9 $a_o$. At this short distance from the nucleus,
the only other contribution to the density is the core contribution, which can be
subtracted out and is unchanged during chemical processes or CO. 
Most relevant 
to the understanding of 
CO-driven transitions and disproportionation is the (actual or relative)
{\it difference} in $3d$ occupations $\Delta n_d$, which
is given directly, {\it without any integration}, by the difference 
in the radial $3d$ densities at their peaks,
where there are no competing orbital occupations to confuse charge counting.
This specifically defined $3d$ occupation differences provides a basis
for building a faithful picture of CO and of characterizing formal valence
differences more realistically.  We consider our computational results\cite{LAPW,wien} for 
a selection of systems, then discuss some of the implications.

\begin{figure}[!htb]
\includegraphics[width=\columnwidth,angle=0]{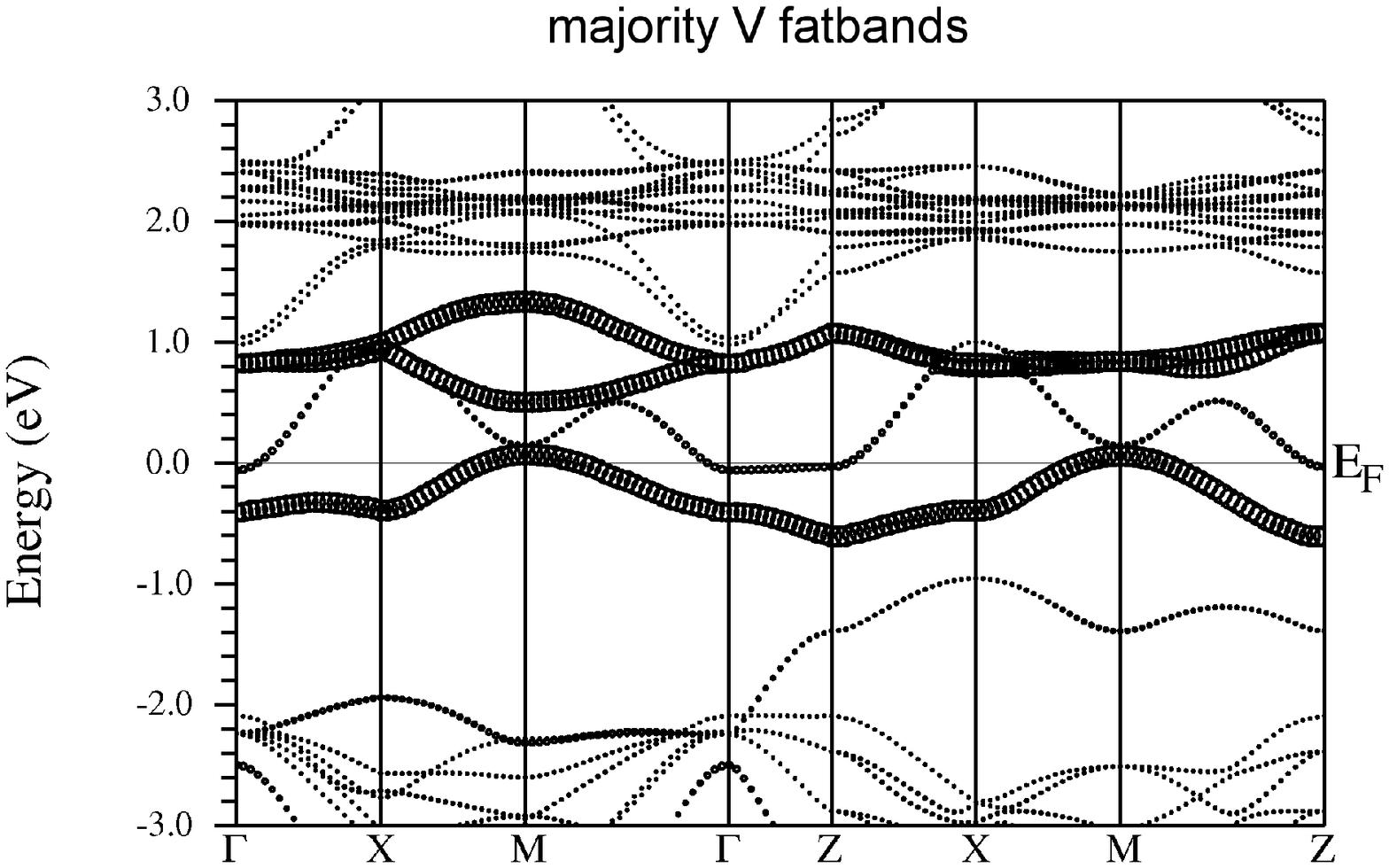}
\includegraphics[width=\columnwidth,angle=0]{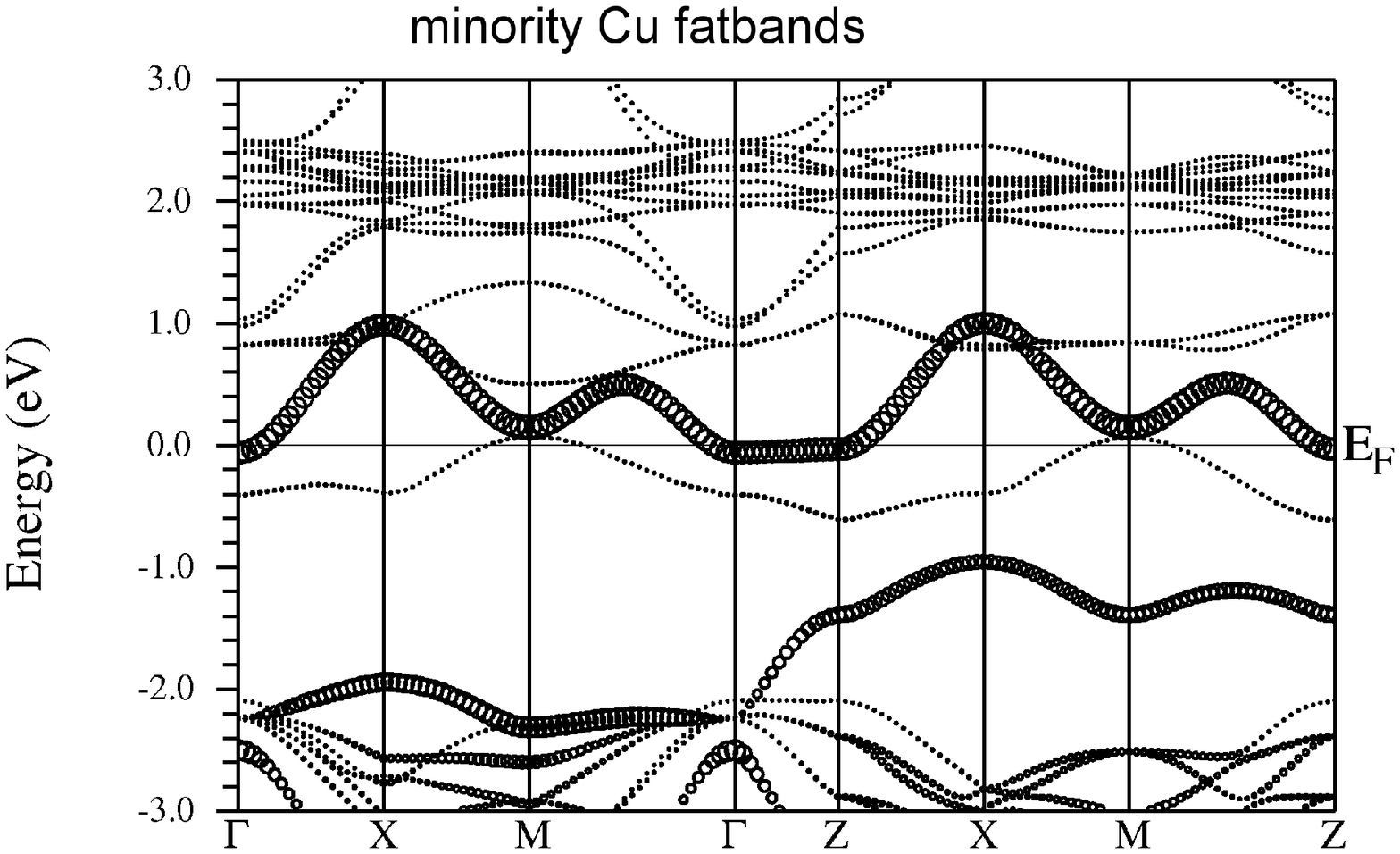}
\caption{(Color online)
Top: bands near the Fermi energy/bandgap in the $d^1 - d^9$ magnetic, nearly
Mott insulating, configuration of La$_2$VCuO$_6$. The $d_{xy}$-up band is
correlation-split off from the other two $t_{2g}$ bands and fully occupied. 
Bottom: the Cu fatbands for the same system, showing one unoccupied
Cu minority $d_{x^2-y^2}$ band correlation-split from the $d_{z^2}$ band.    
The other $d$ bands fall outside this energy range. 
\label{VCuBands}
}
\end{figure}

\underline{La$_2$VCuO$_6$} (LVCO) is a double perovskite compound providing a 
vivid and illustrative example. 
Our earlier study\cite{vcu} revealed
two competing configurations for the ground state. Using conventional identifications, one is the V$^{4+}$ $d^1$,
Cu$^{2+}$ $d^9$ magnetic configuration (with bands shown in 
Fig. \ref{VCuBands}) identified as such because (1) there is one
band of strong V $d$ character occupied and one band of strong Cu $d$ character
unoccupied, and (2) the moments on both V and Cu, 0.7 $\mu_B$, are representative
of many cases of spin-half moments reduced by hybridization with O $2p$ orbitals.  
The other configuration is the nonmagnetic $d^0-d^{10}$ band insulator: all Cu $d$ bands 
are occupied, all V
$d$ bands are unoccupied -- a conventional ionic band insulator in all respects.
The identification of formal valence (or oxidation state) is crystal clear.

\noindent
The radial charge densities of V and of Cu for both configurations
reveal an unsettling feature: the actual $3d$ occupations $n_d$
of each of these V and Cu ions are {\it identical for both configurations}, in spite of the
unit difference in their formal charges. (Identical in this paper means to better than
0.5\% $\sim 0.01 e^-$, in terms of the differences of charge density at their peaks.)
Thus ions with {\it no real difference} in $3d$ occupation can behave as if they comprise
charge states differing by unity.  
Changes in spin-orbital occupations, which quantify spin, orbital, and charge 
differences between the two states, can be quantified by the LDA+U 
spin-orbital occupations. For the V $d^1$ $d_{xy}$ (Jahn-Teller split)
orbital, the majority-minority difference
is 0.70, which accounts for all of the moment. The difference of 0.65 between $d_{xy}$ and 
each of the other $t_{2g}$ characterizes the Jahn-Teller distortion. The increase 
in charge of the $d_{xy}$ orbitals (both spins), 0.55,  compared to the $d^0$ state,  
is absorbed more or less
uniformly from all other (nominally unoccupied) spin orbitals. Similarly for Cu, 
the $d^9$ hole results from a difference of charge in the minority $d_{x^2-y^2}$
orbital of 0.6, with the other hole charge being distributed nearly uniformly over
the other nine (nominally but not actually fully occupied) spin-orbitals. In both cases the moment
arises entirely from the single magnetic orbital as the simple picture
would suggest, while all other orbitals are unpolarized.  
This happens, conspicuously, with
{\it no change} in $n_d$ for either V or Cu. Charge is redistributed to one orbital
from the others, and strongly spin-imbalanced within that orbital.
Even with insulators with ``obvious'' charge states, $3d$ orbital occupations can
range over the values [0,1].

%
We look at additional cases before
addressing some of the implications.

\begin{figure}[!htb]
\begin{center}
\includegraphics[width=\columnwidth,angle=0]{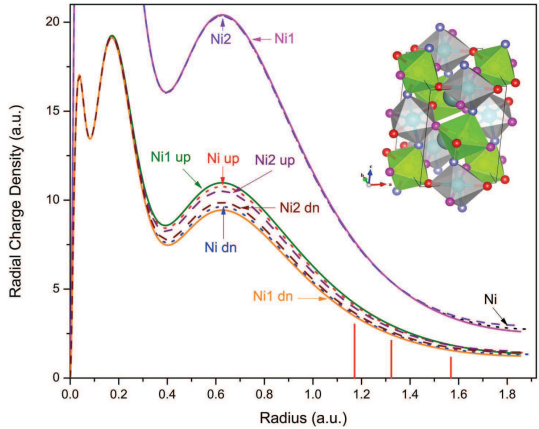}
\end{center}
\caption{(Color online)
Radial charge density (upper curve) of YNiO$_3$ for $Pbnm$ Ni and $P2_1/n$ 
Ni1 and Ni2, showing there is no
difference at the peak, which reflects the $3d$ occupation of the ion; a small
difference shows up near the sphere boundary.  The spin
decompositions give easily visible differences.
The vertical lines at the bottom right indicate conventional Ni$^{4+}$, Ni$^{3+}$,
and Ni$^{2+}$ ionic radii, which have no relation to the (unvarying) $3d$ occupation.
Inset: Structure of the broken symmetry $P2_1/n$ phase, showing the
rotation in the $a-b$ plane and tilting along the $c$ axis of the
NiO$_6$ octahedra (Ni is inside)  and the
($\pi,\pi,\pi$) ordering of the Ni1 and Ni2 octahedra.
\label{Radial}
}
\end{figure}

\underline{Rare earth (${\cal R}$) nickelates ${\cal R}$NiO$_3$} 
display a first order structural and MIT of great current interest. 
The $Pbnm$ (GdFeO$_3$
structure) $\rightarrow$ $P2_1/n$ transformation results in a large Ni1O$_6$ and a
small Ni2O$_6$ octahedron, with Ni-O distances of
2.015$\pm$0.015 \AA~ and 1.915$\pm$0.025 \AA, respectively,
that are not otherwise strongly distorted; see the inset of Fig.~\ref{Radial}.
At a temperature that varies smoothly from 600K to 300K with increasing
${\cal R}$ ionic radius, the resistivity of these nickelates 
drops sharply.\cite{Garcia1992,Torrance1992}
We focus on YNiO$_3$; with its small ionic radius, it is one of the more strongly
distorted members, and the resulting narrowed bandwidths make it more prone
to strong correlation and CO tendencies.\cite{Mazin}
Structural changes at the MIT have been studied 
extensively,\cite{Garcia1992,ynio3,Alonso,Alonso1999,I.Vobornik1999} 
which together with x-ray absorption spectral 
splittings\cite{Staub,Piamonteze,Medarde2009}
have been interpreted in terms of charge disproportionation 
(or CO) 2Ni$^{3+} \rightarrow$ 
Ni$^{3+\delta}$ + Ni$^{3-\delta}$, with $\delta \approx$ 0.3 for YNiO$_3$.\cite{Staub}

This MIT in the nickelates has been recognized as paradigmatic by theorists.  
Mizokawa et al. modeled this system\cite{Miza}  with a multiband Hartree-Fock model in the
charge-transfer regime and found evidence for
CO on the {\it oxygen} sublattice for larger ${\cal R}$ cations, 
but concluded that YNiO$_3$
was representative of a CO transition on the Ni sites.  Mazin {\it et al.}\cite{Mazin} surveyed
the competition between Jahn-Teller distortion of the $d^7$ ion
and CO and also concluded that
YNiO$_3$ is 
a prime example of 
a CO $d^6 + d^8$ system. Lee {\it et al.} have investigated\cite{Balents} 
a two band model for this system with a CO interaction
in mean field,
emphasizing CO effects.  On the other hand,
Yamamoto and Fujiwara\cite{Fujiwara2002} reported a very small ($\sim$0.03 $e^-$)
density functional based 
charge difference.

\noindent
For the assumed (for simplicity) ferromagnetic order the calculated Ni1 and 
Ni2 moments are
1.4 and 0.65 $\mu_B$ respectively for YNiO$_3$ and several other members of this class,
so these values are not sensitive to the magnitude of the distortion.
They coincide with the values 
obtained from neutron diffraction,\cite{ynio3}
1.4(1) and 0.7(1) $\mu_B$ respectively, in the magnetically ordered phase.
It is intriguing that the same moments were obtained in fully relaxed LaNiO$_3$/LaAlO$_3$ 
monolayer superlattices.\cite{Blanca}
 
The $3d$ occupations, obtained as above directly from the maximum in
the radial charge density plots in 
Fig. \ref{Radial}, are identical for Ni1, Ni2, and the single Ni site in the high
temperature phase: there is no $3d$ charge transfer, or disproportionation, across the transition.  
The majority and minority radial densities 
 and integrated charges of course differ (see Fig. \ref{Radial}) 
as they must to give the moment, but the total
$3d$ occupation is inflexible.  
This constancy of the $3d$ occupation across the transition, and 
equality for Ni1 and Ni2, is inconsistent with microscopic disproportionation.

\begin{figure}[!htb]
\begin{center}
\includegraphics[width=\columnwidth,angle=0]{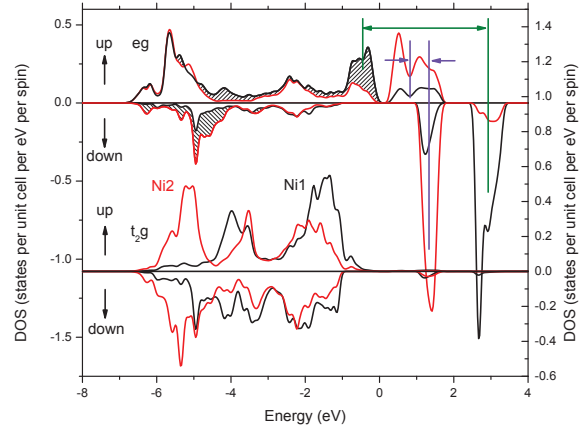}
\end{center}
\caption{(Color online)
Spin-decomposed Ni $t_{2g}$ and $e_g$ density of states for the Ni1 and Ni2
ions in the insulating $P2_1/n$ broken symmetry phase. The hashed regions 
illustrate the spectral origin of the enhanced moment of Ni1 relative to Ni2. 
The horizontal arrows illustrate the large difference in spin splittings, the result of the
combination of Hund's coupling and Coulomb $U$= 5.7 eV. 
\label{DOS}
}
\end{figure}

To illustrate
the spin-orbital spectral density redistribution, the projected densities of states
are shown in Fig. \ref{DOS}.  All $t_{2g}$ states are
filled and irrelevant. 
The $e_g$ spectral distribution is non-intuitive: weight from -5 eV
spin-down is transferred to -1 eV spin-up.  
The majority $e_g$ states just below the gap are strongly Ni1 in
character, while the unoccupied bands just above the gap are primarily Ni2. Such
behavior is expected for different charge states, similarly to the behavior in
LVCO above; however, the total $3d$ occupation is identical.

The main differences between Ni1 and Ni2 
show up  in the {\it unoccupied} $e_g$ states: the Ni1 spin splitting is 3.5 eV, a 
reflection of the on-site repulsion that opens the Mott gap in the majority $e_g$
states, rather than Hund's exchange splitting. The origin of the Ni2 moment is murky, not
identifiable with any occupied spectral density peak. Note that in a Ni$^{2+}$ $+$ 
Ni$^{4+}$ CO picture, Ni2 would be nonmagnetic.  Not only is this calculated behavior not consistent
with a CO picture, it involves redistribution not accounted for in
any simple model. In spite of identical $3d$ charges, the Ni1 and Ni2 core energies
differ by up to 1.5 eV.

\noindent
\underline{CaFeO$_3$}, another perovskite that displays the same 
$Pbnm \rightarrow P2_1/n$ structural
change at T$_{MI}$ as the nickelates,
is also explained\cite{Takano} in CO language that invokes the unusually
high (penta)valent state Fe$^{5+}$.
Analogously to YNiO$_3$, we obtain identical $3d$ occupations
for Fe1 and Fe2 ions. Quantum chemical embedded cluster 
calculations\cite{QuantumChemical} and LDA+U studies\cite{LDA+U,saha,matsuno} 
had noted that the
Fe charge in both ``disproportionated'' sites differed little, but neither
quantified the occupation as we have for YNiO$_3$ and CaFeO$_3$.  
The pentavalent state of Fe has most often been identified from M\"ossbauer isomer shift
data, but Sadoc {\it et al.}\cite{QuantumChemical} concluded the difference in isomer shift is
primarily a measure of the covalency (Fe-O distance) rather than any real charge on Fe.

\underline{AgNiO$_2$}, a triangular, magnetically frustrated lattice compound with
nominal Ni$^{3+}$ ions, undergoes a structural transition at 365 K although remaining
metallic.\cite{wawrzy1,wawrzy2,chung,pascut}  Three inequivalent Ni sites arise, 
with a high spin Ni1 ion in an enlarged
octahedron and two low spin Ni2, Ni3 = Ni2,3 ions in small octahedra.  Based on the
structural changes (which were quantified in terms of bond valence sums), the magnetic
moments, and resonant x-ray scattering that confirms a calculated $\sim$1 eV difference in core
level energies between Ni1 and Ni2,3, this transition has been welcomed 
as the first realization
of such a highly unusual 3$e_g^1 \rightarrow e_g^2 +2 e_g^{0.5}$ 
type of CO.  Furthermore, using
the charge difference per unit core level splitting of 0.66 $e$/eV led to an
inferred charge disproportionation of $\sim$1.65$e$, {\it i.e.} Ni1$^{2+}$ +
2 Ni2,3$^{3.5+}$. 
We have reproduced several of the first principles results\cite{wawrzy1,pascut}
that were used to support CO.
The calculations
give a large moment ($> 1 \mu_B$) on high-spin Ni1 and very weak moments 
($\sim$0.1 $\mu_B$) on low-spin Ni2,3 ions. 
We find, as in the cases above, that $n_d$ for the three sites are {\it identical}.
Moreover, our calculated 
core level differences, 0.6-0.8 eV, are roughly consistent with reported
values\cite{pascut} ($\sim$1 eV). 

\underline{V$_4$O$_7$} represents another oxide currently explained by a CO-driven MIT.
It is structurally more involved, but
first principles calculations of moments and geometries again have produced several
results corroborating the experimental data\cite{v4o7,Hodeau} and were used to support
CO into V$^{3+}$ and V$^{4+}$ charge states on specific sites.
As in the instances above, we find no differences
in $n_d$:
the occupations are indistinguishable.
The site energy differences, measured by differences in $1s, 2s, 2p$ core
levels, differ by 0.9-1.2 eV for two sites, similar
to the nickelates. The interplay of orbital order, structural distortions, and
possible spin-singlet formation of half of the V ions provide a rich array of
degrees of freedom, which can operate without need for disproportionation. 

\underline{Implications.} We have established that, for several instances of CO
transition systems as well as for the two self-evident 
charge states of LVCO, there is no difference
in the $3d$ occupations for the different ``charge states''  
that have been used to categorize their behavior. Such identification is possible
because a choice of a region for integration is avoided; the peak charge region rather 
than tails of orbitals are used in the identification.  
This finding of constancy sharpens several reports of ``small charge differences'' 
between differing charge states
({\it viz.} Luo {\it et al.}\cite{Luo}
for doped manganites; Haldane and Anderson\cite{haldane} in a multi-orbital 
Anderson model, and 
Raebiger {\it et al.}\cite{lany} from DFT calculations for TM impurities in semiconductors;
Yamamoto and Fujiwara\cite{Fujiwara2002} and also 
Park {\it et al.}\cite{rnio_millis} for nickelates). 

We see two primary implications: (1) the conceptual basis underlying a substantial
aspect of transition metal physics is misleading, and (2) modeling of structural and
electronic transitions has, at least in several conspicuous cases, incorporated the wrong 
mechanisms by invoking inactive degrees of freedom. Actual cases of CO very likely do exist,
but the burden of proof has shifted.


For these CO systems, the constancy of $n_d$ suggests that $U_d$ is too large to allow change
in occupation $n_d$ in or near the ground state (in the 
cases we discuss, and similar ones). 
In insulators the charge is more physically pictured in terms
of (fully occupied) Wannier functions (WFs) than in terms of ambiguous 
populations of atomic orbitals, making them appear to be inviting. 
However, WFs are far from unique and, like molecular orbitals, WFs 
contain charge that cannot objectively be assigned to one atom or another,
so a WF viewpoint is not promising. 

A broader implication is that modeling of coupled structural 
and electronic transitions in
terms of charges\cite{Mazin,Balents} from atomic-like orbitals must be treated with 
caution: 
charge fluctuations in these systems are too high in energy
to comprise a relevant degree of freedom.
The important energy
differences are characterized in terms of differences in hopping
amplitudes, anion-cation distances, and (not recognized in most
models) resulting changes in site energies, as well as very 
important Hund's rule energies. Models that try to
parametrize (for example) Ni1-Ni2 differences by on-site charge
will not be treating the relevant microscopic degrees of freedom. CO on the
oxygen sublattice\cite{mizokawa,Mazin} may also be problemmatic.

Charge states of ions serve to specify the occupations of spin-orbitals.
The essential degrees of freedom in determining this popular characterization,
which professes to be quantitative, are the spin-orbital occupations, not
as determined from the (real) density matrix but rather from the
site symmetry, crystal symmetry, and the local moment.
The LVCO example illustrates vividly how two different charge states,
for both {\it highly charged} V and {\it moderately charged} Cu, can be
represented by integer occupation of different numbers of orbitals while there
is no change in $n_d$. ``Charge state'' projects onto integrally occupied
orbitals, while the distribution of real charge is strongly non-integral and 
often non-intuitive. These projections are backed up by the number of occupied 
spin-polarized bands (an integer),
by the (discrete) local
symmetry (JT distortion), by the local moment (with its quantization
smeared by hybridization), and by the atomic radii, 
but each one of these characterizations is extremely flexible with a given amount of
$3d$ charge.

More specifically to CO systems, the ionic environment in the high symmetry 
phase requires closer
scrutiny.  In both the nickelates and in V$_4$O$_7$ there is 
evidence of distinct metal
sites {\it above} the transition, in the (on average) symmetric phase, and the structural
similarities of CaFeO$_3$ to $R$NiO$_3$ suggest similar behavior there. 
For nickelates, x-ray absorption
spectra\cite{Medarde2009,Piamonteze} reveal that local signatures of Ni1 and Ni2 sites persist
continuously across the MIT, and both sites also remain when driven across the
phase boundary by pressure.\cite{Ramos} As we have shown, the coordination alone 
({\it i.e.} with identical $n_d$) accounts for on-site
energy differences of $\sim$1 eV in spectra that have often been used to support
disproportionation. The MITs in some of these materials may be primarily order-disorder
type; the onset of long-range order in nickelates results in carrier localization
and gap formation, ergo a MIT but one unrelated to CO. 


We propose therefore that ``charge order'' should be used as the name,  
hence the interpretation,
of a phase transition only if an objective, relevant charge difference is the likely
mechanism; otherwise, the
underlying mechanisms should be identified.
Formal developments may be useful;
for example,  Jiang {\it et al.} have provided a
specification\cite{jiang} of integer charges in an insulator that they propose as
oxidation states (which are identical to charge states in metal oxides.)  
Based on integration over a configuration
space path of the dynamic Born effective charge, their expression 
assigns (in principle) an integer
charge to each atom in any insulator. 
Notably, their specification does not refer to $3d$ charge explicitly and furthermore depends
explicitly on dynamical effects (electron response to ion motion).
Also, many CO interpretations only hold water if the supposed charge
difference $\pm \delta$ is much smaller than unity ($\delta \sim$ 0.3 for
the nickelates).
More experience will be needed to learn how best to
interpret their definition. 


Work at UC Davis was supported by DOE grant DE-FG02-04ER46111.
V.P. acknowledges support from the Spanish Government through 
the Ram\'{o}n y Cajal Program.


\begin{thebibliography}{50}
\bibitem{LAPW} The density functional based linearized augmented plane wave code
{\sc wien2k}\cite{wien} has been used in our
calculations, where we have applied correlation
corrections (LDA+U) as is usually necessary to describe $3d$ oxides. Computational
methods are as described in Ref. \cite{vcu}.
\end{thebibliography}

\end{document}